\documentclass[aip,reprint]{revtex4-1}
\usepackage{hyperref}
\usepackage{graphicx}
\usepackage{epstopdf}

\draft % marks overfull lines with a black rule on the right

\begin{document}

%\preprint{}

\title{Influence of Fe buffer thickness on the crystalline quality and the transport properties of Fe/Ba(Fe$_{1-x}$Co$_x$)$_2$As$_2$ bilayers} %Title of paper



\author{K. Iida}
\email[Electronical address:\,]{k.iida@ifw-dresden.de}
\author{S.\,Haindl}
\author{T.\,Thersleff}
\author{J.\,H\"{a}nisch}
\author{F.\,Kurth}
\author{M.\,Kidszun}
\author{R.\,H\"{u}hne}
\author{I.\,M\"{o}nch}
\author{L.\,Schultz}
\author{B.\,Holzapfel}
%\thanks{}
\affiliation{IFW Dresden, P.\,O.\,Box 270116, 01171 Dresden, Germany}
\author{R.\,Heller}
\affiliation{Institute of Ion Beam Physics and Materials Research, FZ Dresden-Rossendorf, Bautzner Landstr. 400, 01328 Dresden, Germany}

\date{\today}

\begin{abstract}
The implementation of an Fe buffer layer is a promising way to obtain epitaxial growth of Co-doped BaFe$_2$As$_2$ (Ba-122). However, the crystalline quality and the superconducting properties of Co-doped Ba-122 are influenced by the Fe buffer layer thickness, $d_{\rm Fe}$. The well-textured growth of the Fe/Ba-122 bilayer with $d_{\rm Fe}=15\,{\rm nm}$ results in a high $J_{\rm c}$ of $0.45$\,MAcm$^{-2}$ at 12\,K in self-field, whereas a low $J_{\rm c}$ value of 61000\,Acm$^{-2}$ is recorded for the bilayer with $d_{\rm Fe}=4\,{\rm nm}$ at the corresponding reduced temperature due to the presence of grain boundaries.
\end{abstract}


\maketitle
Strong bonding and small misfit between functional layers and substrates are preferable for epitaxy. Our recent publication demonstrates that implementing an Fe buffer layer is an advantageous way to obtain epitaxial growth of Co-doped BaFe$_2$As$_2$ (Ba-122), since the metallic bond between the Fe and the Co-doped Ba-122 layer takes place at the Fe sublayer within the FeAs tetrahedron.\cite{47} Furthermore, the lattice parameter $a$ of Fe is close to the Fe-Fe distance in the Co-doped Ba-122 unit cell, resulting in a small lattice mismatch of 2\,\%. Here, the (001) surface plane of Fe is rotated 45$^\circ$ in-plane to the iron sublayer in the Co-doped Ba-122 unit cell. As a result, Co-doped Ba-122 films are of excellent crystalline quality and show a high superconducting transition temperature, $T_{\rm c}$, up to 24.6\,K.

Epitaxial Co-doped Ba-122 films with sharp out-of-plane and in-plane texture as well as a high $T_{\rm c}$ of 24.5\,K have been deposited on bare SrTiO$_3$ (STO) substrates.\cite{11} Furthermore, structurally comparable films on bare  (La$_{0.3}$Sr$_{0.7}$)(Al$_{0.65}$Ta$_{0.35}$)O$_3$ (LSAT) with a $T_{\rm c}$ between 22.3 and 22.8\,K have also been realized.\cite{24,39} Detailed transmission electron microscope (TEM) investigations revealed that an interfacial layer of textured Fe was quite frequently observed between epitaxially-grown Co-doped Ba-122 and bare LSAT substrates.\cite{24} Textured Fe was also confirmed by pole figure measurements on different substrates such as STO and LaAlO$_3$,\cite{51} presumably growing at the interface. Hence, Fe may work as a generic buffer layer for epitaxial growth of Co-doped Ba-122 layers on multiple substrates. The Fe layer should be thick enough to cover the substrates fully. Whenever a discontinuous textured Fe layer is grown at the interface, misoriented grains of Co-doped Ba-122 are likely observed, reducing the critical current. Indeed the current-limiting effect of grain boundaries even with small misorientation angles for Co-doped Ba-122 has been reported.\cite{09} Furthermore, the implementation of a thick Fe buffer layer may also work as a strain absorber and a high $J_{\rm c}$ of strain-relaxed Co-doped Ba-122 films can be expected. That might explain why Co-doped Ba-122 films on bare LSAT showed a self-field $J_{\rm c}$ of $\sim10^4$Acm$^{-2}$ in a previous publication,\cite{24} one order of magnitude lower than single crystals,\cite{14} and two orders of magnitude lower than recently reported films,\cite{39,10,43} two of which confirmed the presence of $c$-axis correlated defects.\cite{10,43} Hence one can expect high-$J_{\rm c}$ Co-doped Ba-122 films grown on a perfectly textured Fe layer, which may depend on Fe layer thickness significantly.

In this letter, we report on the influence of the Fe buffer structure by employing two different Fe layer thicknesses ($d_{\rm Fe}$=4\,nm and 15\,nm) on structure and transport properties of the Fe/Ba-122 bilayer system.

Fe layers were deposited with 1250 and 5000 pulses at room temperature on MgO (001) single crystalline substrates by pulsed laser deposition, PLD, using a KrF excimer laser (248\,nm) at a frequency of 10\,Hz in a UHV chamber (base pressure 10$^{-8}$\,mbar). Subsequently, the Fe-covered MgO substrates were heated to 700\,$^\circ$C, which in turn  serves for the deposition of Co-doped Ba-122 layers with a pulse number of 13500. Here, a phase pure target of the nominal composition, BaFe$_{1.8}$Co$_{0.2}$As$_2$, was employed for PLD. Films composition were almost identical to the target composition confirmed by Rutherford Backscattering Spectroscopy (RBS). The detailed PLD target preparation can be found in Ref.\cite{11} The energy density of the laser on the target was 3--5\,Jcm$^{-2}$, and the distance between target and substrate was approximately 7\,cm. 

\begin{figure}
	\centering
		\includegraphics[height=6.5cm,width=8cm]{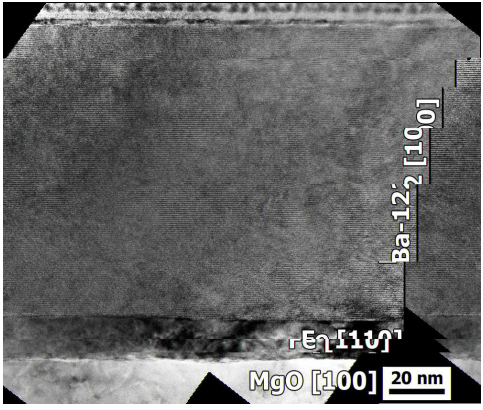}
		\caption{(Color online) The representative microstructure of the well-textured Fe/Ba-122 bilayer ($d_{\rm Fe}$=15\,nm). The Co-doped Ba-122 layer is clean and no appreciable defects are observed. The figure has been digitally stitched together from three separate images acquired under the same conditions in order to present a larger area of the film.} 
\label{fig:figure1}
\end{figure}

The representative microstructure of the Fe/Ba-122 bilayer with 5000 pulses of Fe is presented in fig.\,\ref{fig:figure1} as a cross-sectional bright-field TEM micrograph, taken in a ${\rm C_s}$-corrected 300\,kV, FEI Titan. No misoriented grains were observed throughout the entire lamella and the Co-doped Ba-122 layer was clean. However, a few line defects per 100\,$\mu$m were observed in the Co-doped Ba-122 layer (not shown in this letter). Nevertheless, such a low density of defects shows no influence on the angular-dependent $J_{\rm c}$ described later. The average thickness of the Fe and the Ba-122 layers was around 15 and 100 nm, respectively, which was in good agreement with the RBS analyses. The respective growth rates of Fe and Co-doped Ba-122 are estimated to 0.026\,\AA/pulse and 0.074\,\AA/pulse. Based on this, $d_{\rm Fe}$ with 1250 pulses is estimated to $d_{\rm Fe}\approx4$\,nm.

\begin{figure}
	\centering
		\includegraphics[height=5cm,width=6.5cm]{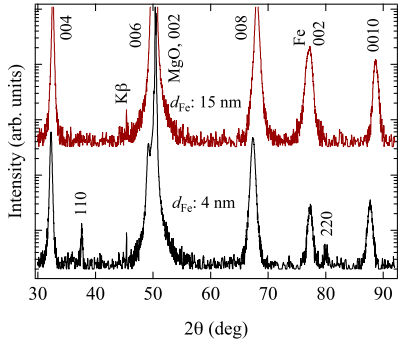}
		\caption{(Color online) $\theta\rm/2\theta$-\,scans of Fe/Ba-122 bilayers with $d_{\rm Fe}$=4\,nm and 15\,\rm nm on (001) MgO substrates in Bragg-Brentano geometry using Co-K$\alpha$ radiation. No secondary phase was observed, proving a high phase purity. However, additional 110 and 220 peaks of Co-doped Ba-122 were observed for the bilayer with $d_{\rm Fe}=4\,\rm nm$.} 
\label{fig:figure2}
\end{figure}

The out-of-plane texture and phase purity of Fe/Ba-122 bilayers were examined by $\theta\rm/2\theta$-\,scans using Co-K$\alpha$ radiation. The crystalline quality were evaluated via $\omega$\,-\,scans of the 004 reflection, as well as 103 $\phi$\,-\,scans for Co-doped Ba-122 and 110 $\phi$\,-\,scans for Fe using Cu-K$\alpha$ radiation. Shown in fig.\,\ref{fig:figure2}\, are the $\theta\rm/2\theta$-\,scans for the Fe/Ba-122 bilayers with different $d_{\rm Fe}$. Only 00$l$ peaks of the Co-doped Ba-122 together with the 002 reflection of Fe and MgO were observed for $d_{\rm Fe}=15$\,nm. However, additional 110 and 220 peaks of Co-doped Ba-122 were observed for the bilayer with $d_{\rm Fe}=4\,\rm nm$, which is due to the Volmer-Weber growth of Fe,\cite{47} leading to tilted grains of Co-doped Ba-122. Such a structural imperfection can be modified by increasing nominal $d_{\rm Fe}$, since the surface of the MgO substrate is not fully covered with Fe. Pole figure measurements showed that the full epitaxy of Co-doped Ba-122 was only confirmed for $d_{\rm Fe}=15$\,nm. Here the epitaxial relation is (001)[100]Ba-122$\|$(001)[110]Fe$\|$(001)[100]MgO. The full width at half maximum (FWHM), $\Delta\omega$, of the 004 rocking curve for the film with $d_{\rm Fe}$=15\,nm was $0.64^\circ$, whereas a large $\Delta\omega$ of $1.33^\circ$ was observed for the film with $d_{\rm Fe}$=4\,nm. The average $\Delta\phi$ of the 103 reflection of Co-doped Ba-122 on $d_{\rm Fe}$=4\,nm showed with $\Delta\phi=1.56^\circ$ a larger FWHM than the Fe layer itself ($\Delta\phi=1.10^\circ$) in contrast to the film with a thicker Fe layer, which showed a lower value of $\Delta\phi=0.95^\circ$ compared to $\Delta\phi=1.05^\circ$ of Fe.

The bilayers were photolithographically patterned and ion beam etched to form bridges of 100--500\,$\mu$m width and 0.8\,mm length for transport measurements. Gold electrodes were deposited by PLD at room temperature, to which Cu wires of 100\,$\mu$m diameter were connected by silver paint, ensuring a low contact resistance. Superconducting properties were measured in a Physical Property Measurement System (PPMS, Quantum Design) by a standard four-probe method with a criterion of 1\,$\rm\mu Vcm^{-1}$ for evaluating $J_{\rm c}$. In the angular-dependent $J_{\rm c}$ measurements, the magnetic field, $H$, was applied in the maximum Lorentz force configuration ($H$ perpendicular to $J$) at an angle $\Theta$ measured from the $c$-axis. A small DC of 10\,$\mu$A ($J$=15--77\,Acm$^{-2}$) was employed for resistivity measurements, $\rho$($T$). $T_{\rm c}$ is defined as $90\%$ of the normal resistivity at 30\,K.

\begin{figure}
	\centering
		\includegraphics[height=4cm,width=8cm]{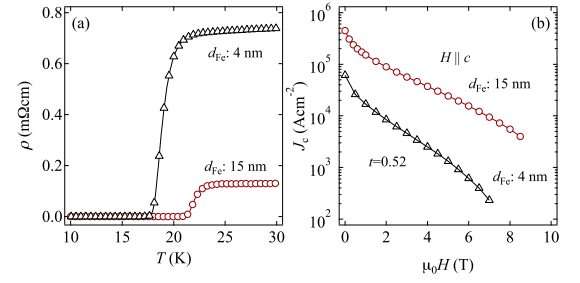}
		\caption{(Color online) (a) Resistivity traces for the Fe/Ba-122 bilayers with different $d_{\rm Fe}$. The respective $T_{\rm c}$ of Fe/Ba-122 with $d_{\rm Fe}=$15\,nm and 4\,nm were 23.1\,K and 20.2\,K. (b) Transport $J_{\rm c}$ for $H$$\parallel$${c}$ of the bilayers with $d_{\rm Fe}=4\,\rm nm$ and 15\,nm at a reduced temperature of $t=T/T_{\rm c}=0.52$.} 
\label{fig:figure3}
\end{figure}

$T_{\rm c}$ of Fe/Ba-122 is increased from 20.2\,K to 23.1\,K with increasing $d_{\rm Fe}$ (fig.\,\ref{fig:figure3}\,(a)). There are two plausible reasons for variation in $T_{\rm c}$ such as microstructure and proximity effects at the Fe/Ba-122 boundary. As stated earlier, the Fe/Ba-122 with thin Fe buffer contains a lot of grain boundaries, leading to oxidization along the grain boundaries significantly, which may reduce $T_{\rm c}$. Grain boundaries may also increase the normal state resistivity, since electrons are scattered at grain boundaries.  

The current-limiting effect of grain boundaries can be seen in $J_{\rm c}-H$ characteristics (fig.\,\ref{fig:figure3}\,(b)). In comparison to the bilayer with $d_{\rm Fe}=15\,\rm nm$, transport $J_{\rm c}$ of the bilayer with $d_{\rm Fe}=4\,\rm nm$ shows lower values at the entire magnetic field range. On the other hand, the bilayer with $d_{\rm Fe}=15\,\rm nm$ showed a high self-field $J_{\rm c}$ of 0.45\,MAcm$^{-2}$ at 12\,K, which is almost 40 times higher than the value of the Co-doped Ba-122 on bare LSAT in our previous publication.\cite{24}

\begin{figure}
	\centering
			\includegraphics[height=6.5cm,width=8cm]{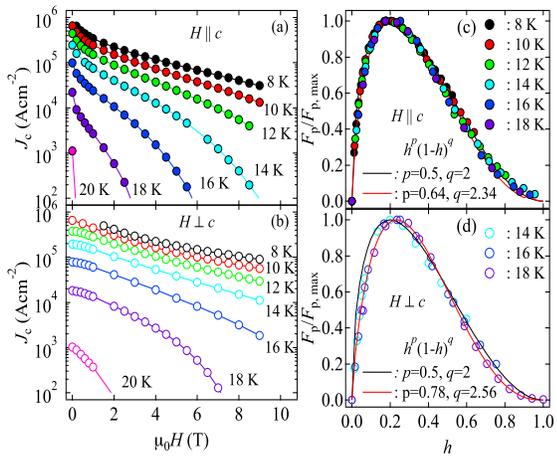}
		\caption{(Color online) $J_{\rm c}-H$ characteristics of the Fe/Ba-122 bilayer ($d_{\rm Fe}=15\,\rm nm$) at different temperatures. The field was applied parallel (a) and perpendicular to the $c$-axis (b). Normalized pinning force, $F_{\rm p}/F_{\rm p,max}$, as a function of reduced field for (c) $H$$\parallel$${c}$ and (d) $H$$\perp$${c}$. The black lines in both figures are calculated with the Kramer model ($p=0.5, q=2$) for the aim of comparison.} 
\label{fig:figure4}
\end{figure}

Transport $J_{\rm c}$ of the Fe/Ba-122 bilayer with $d_{\rm Fe}=15\,\rm nm$ as a function of magnetic field at different temperatures has been evaluated for both major directions (i.e. $H$$\parallel$${c}$ and $H$$\perp$${c}$). It is clear from fig.\,\ref{fig:figure4} (a) and (b) that $J_{\rm c}$ values for $H$$\parallel$${c}$ are always lower than for $H$$\perp$${c}$, indicating that the flux pinning is anisotropic.

The scaling behavior of the pinning force density for $H$\,$\parallel$\,${c}$ is shown in fig.\,\ref{fig:figure4}\,(c). Here, the reduced field, $h$, is defined as $h=H/H_{\rm irr}$, where $H_{\rm irr}$ is the irreversibility field. Below 14\,K, $H_{\rm irr}$ has been evaluated by extrapolation according to the Kramer method. All data of the normalized pinning force, $F_{\rm p}/F_{\rm p,max}$, fall onto a master curve of $h^{p}(1-h)^q$ with the exponents $p=0.64, q=2.34$ regardless of temperature, which is explained by the Kramer model for shear breaking of the flux line lattice as main reason for depinning.\cite{32} It is plausible that the elastic constant of flux line lattice shear, $C_{66}$, for Co-doped Ba-122 is very small due to the large Ginzburg-Landau parameter, $\kappa=\lambda/\xi\approx 70$.\cite{33}
In contrast to $H$\,$\parallel$\,${c}$, the pinning force for $H$$\perp$${c}$ can only be scaled in a temperature range between 14\,K to 18\,K with $h^{0.78}(1-h)^{2.56}$, whereas the scaling is not possible below 12\,K, which implies a different pinning mechanism in this temperature region. 

\begin{figure}
	\centering
		\includegraphics[height=4cm,width=8cm]{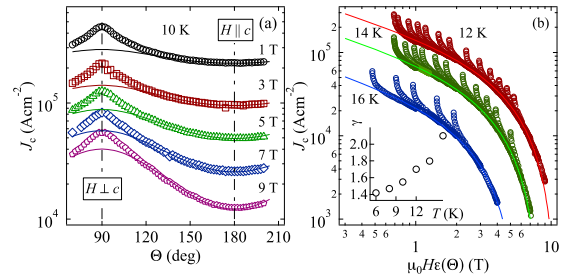}
		\caption{(Color online) (a) $J_{\rm c}(\Theta)$ of the Fe/Ba-122 bilayer ($d_{\rm Fe}=15\,\rm nm$) measured under several magnetic fields at 10\,K . No additional peaks except at $H$\,$\perp$\,${c}$ were observed. The solid lines represent the random pinning contribution. (b) All $J_{\rm c}$ data can be scaled with an anisotropy parameter of $\gamma$ ranging 1.4--2.1 at given temperatures except for the angular range close to $H$\,$\perp$\,${c}$. The re-calculated $J_{\rm c}$ from pinning force scaling described in fig.\,\ref{fig:figure4} was also plotted (solid lines). The $\gamma$ values were observed to decrease with decreasing temperature, as shown in the inset, which shows the same behavior as Co-doped Ba-122 on bare LSAT. (See Ref. \cite{24}) This temperature dependence of $\gamma$ is an evidence for multi-band superconductivity.} 
\label{fig:figure5}
\end{figure}

Shown in fig.\,\ref{fig:figure5}(a) are $J_{\rm c}(\Theta)$ data measured under different magnetic fields at 10\,K. $J_{\rm c}(\Theta)$ always has a broad maximum positioned at $\Theta=90^{\circ}$ ($H$\,$\perp$\,${c}$) mainly due to the intrinsic pinning arising from coupled $ab$-planes. It should be noted that this peak is also observed in a Fe/Ba-122 bilayer with a 300\,nm thick superconducting layer. Albeit the high $J_{\rm c}$ of over $0.45$\,MAcm$^{-2}$ for $H$$\perp$${c}$ at 1\,T, the corresponding value at $H$$\parallel$${c}$ is remarkably reduced by almost half. Such a reduction in $J_{\rm c}$ becomes more significant with applied fields due to a rather low density of line defects along the $c$-axis, which is reflected in the TEM observation, as shown in fig.\,\ref{fig:figure1}, and hence a low $H_{\rm irr}$ in this direction.

For a more detailed investigation, the anisotropic Ginzburg-Landau scaling is applied to $J_{\rm c}(\Theta)$ data in order to distinguish between correlated and uncorrelated defects, i.e. $J_{\rm c}(\Theta)$ is plotted as a function of an effective magnetic field, $H_{\rm eff}$, where $H_{\rm eff}=H\epsilon(\Theta)$, $\epsilon(\Theta)=\sqrt{\cos ^2(\Theta)+\gamma^{-2}\sin ^2(\Theta)}$, and $\gamma$ is the mass anisotropy ratio.\cite{40} The scaling behavior of $J_{\rm c}(\Theta)$ as a function of $H_{\rm eff}$ at representative temperatures is displayed in fig.\,\ref{fig:figure5}\,(b). All data at one temperature except in the vicinity of $H$$\perp$$c$ fall onto a single curve with relatively small anisotropies, $\gamma$, of 1.4--2.1, indicating that random defects are dominant in this angular region.

To conclude, the crystalline quality and the superconducting properties of the Co-doped Ba-122 films on Fe-buffered (001) MgO have been characterized. In contrast to the Fe/Ba-122 bilayer with 4\,nm Fe, higher $T_{\rm c}$ and sharper out-of-plane and in-plane texture of the Fe/Ba-122 bilayers can be obtained for 15\,nm thick Fe. The well-textured Fe/Ba-122 bilayer on MgO with $d_{\rm Fe}=15\,\rm nm$ shows a superior $J_{\rm c}-H$ characteristic at lower field regime. Normalized pinning force curves are well fitted by the predictions of the Kramer model for shear breaking. Angular-dependent $J_{\rm c}$ measurements exhibit no $c$-axis peak, whereas a broad maximum positioned at $\Theta=90^{\circ}$ ($H$\,$\perp$\,${c}$) is observed regardless of temperature and field, which is due to the intrinsic pinning arising from correlated $ab$-planes. Finally, further improvement of $J_{\rm c}-H-\Theta$  performances will be possible by introducing artificial pinning centers, since the present films have a clean microstructure.

\begin{acknowledgments}
The authors thank G.\,Fuchs, S.\,Trommler, J.\,Engelmann and R.\,Prozorov for fruitful discussion, M.\,Sparing and R.\,G\"{a}rtner for help with the gold electrode preparation as well as  J.\,Scheiter for help with FIB cut samples. We are also grateful to M.\,K\"{u}hnel and U.\,Besold for their technical support. This work was partially supported by the EU Marie-Curie RTN NESPA and the German Research Foundation, DFG.
\end{acknowledgments}


\begin{thebibliography}{10}
\makeatletter
\providecommand \@ifxundefined [1]{%
 \ifx #1\undefined \expandafter \@firstoftwo
 \else \expandafter \@secondoftwo
\fi
}%
\providecommand \@ifnum [1]{%
 \ifnum #1\expandafter \@firstoftwo
 \else \expandafter \@secondoftwo
\fi
}%
\providecommand \enquote [1]{``#1''}%
\providecommand \bibnamefont  [1]{#1}%
\providecommand \bibfnamefont [1]{#1}%
\providecommand \citenamefont [1]{#1}%
\providecommand\href[0]{\@sanitize\@href}%
\providecommand\@href[1]{\endgroup\@@startlink{#1}\endgroup\@@href}%
\providecommand\@@href[1]{#1\@@endlink}%
\providecommand \@sanitize [0]{\begingroup\catcode`\&12\catcode`\#12\relax}%
\@ifxundefined \pdfoutput {\@firstoftwo}{%
 \@ifnum{\z@=\pdfoutput}{\@firstoftwo}{\@secondoftwo}%
}{%
 \providecommand\@@startlink[1]{\leavevmode}%
 \providecommand\@@endlink[0]{}%
}{%
 \providecommand\@@startlink[1]{%
  \leavevmode
  \pdfstartlink
   attr{/Border[0 0 1 ]/H/I/C[0 1 1]}%
   user{/Subtype/Link/A<</Type/Action/S/URI/URI(#1)>>}%
  \relax
 }%
 \providecommand\@@endlink[0]{\pdfendlink}%
}%
\providecommand \url  [0]{\begingroup\@sanitize \@url }%
\providecommand \@url [1]{\endgroup\@href {#1}{\urlprefix}}%
\providecommand \urlprefix [0]{URL }%
\providecommand \Eprint[0]{\href }%
\@ifxundefined \urlstyle {%
  \providecommand \doi [1]{doi:\discretionary{}{}{}#1}%
}{%
  \providecommand \doi [0]{doi:\discretionary{}{}{}\begingroup
  \urlstyle{rm}\Url }%
}%
\providecommand \doibase [0]{http://dx.doi.org/}%
\providecommand \Doi[1]{\href{\doibase#1}}%
\providecommand \selectlanguage [0]{\@gobble}%
\providecommand \bibinfo [0]{\@secondoftwo}%
\providecommand \bibfield [0]{\@secondoftwo}%
\providecommand \translation [1]{[#1]}%
\providecommand \BibitemOpen[0]{}%
\providecommand \bibitemStop [0]{}%
\providecommand \bibitemNoStop [0]{.\EOS\space}%
\providecommand \EOS [0]{\spacefactor3000\relax}%
\providecommand \BibitemShut [1]{\csname bibitem#1\endcsname}%
%</preamble>
\bibitem{47}%
  \BibitemOpen
  \bibfield{author}{%
  \bibinfo {author} {\bibfnamefont{T.}~\bibnamefont{Thersleff}}, \bibinfo
  {author} {\bibfnamefont{K.}~\bibnamefont{Iida}}, \bibinfo {author}
  {\bibfnamefont{S.}~\bibnamefont{Haindl}}, \bibinfo {author}
  {\bibfnamefont{M.}~\bibnamefont{Kidszun}}, \bibinfo {author}
  {\bibfnamefont{D.}~\bibnamefont{Pohl}}, \bibinfo {author}
  {\bibfnamefont{A.}~\bibnamefont{Hartmann}}, \bibinfo {author}
  {\bibfnamefont{F.}~\bibnamefont{Kurth}}, \bibinfo {author}
  {\bibfnamefont{J.}~\bibnamefont{H{\"a}nisch}}, \bibinfo {author}
  {\bibfnamefont{R.}~\bibnamefont{H{\"u}hne}}, \bibinfo {author}
  {\bibfnamefont{B.}~\bibnamefont{Rellinghaus}}, \bibinfo {author}
  {\bibfnamefont{L.}~\bibnamefont{Schultz}},\ and\ \bibinfo {author}
  {\bibfnamefont{B.}~\bibnamefont{Holzapfel}},\ }%
  \bibfield{journal}{%
  \bibinfo {journal} {Appl. Phys. Lett.}\ }%
  \textbf{\bibinfo {volume} {97}},\ \bibinfo {pages} {022506} (\bibinfo {year}
  {2010})\BibitemShut{NoStop}%
\bibitem{11}%
  \BibitemOpen
  \bibfield{author}{%
  \bibinfo {author} {\bibfnamefont{K.}~\bibnamefont{Iida}}, \bibinfo {author}
  {\bibfnamefont{J.}~\bibnamefont{H{\"a}nisch}}, \bibinfo {author}
  {\bibfnamefont{R.}~\bibnamefont{H{\"u}hne}}, \bibinfo {author}
  {\bibfnamefont{F.}~\bibnamefont{Kurth}}, \bibinfo {author}
  {\bibfnamefont{M.}~\bibnamefont{Kidszun}}, \bibinfo {author}
  {\bibfnamefont{S.}~\bibnamefont{Haindl}}, \bibinfo {author}
  {\bibfnamefont{J.}~\bibnamefont{Werner}}, \bibinfo {author}
  {\bibfnamefont{L.}~\bibnamefont{Schultz}},\ and\ \bibinfo {author}
  {\bibfnamefont{B.}~\bibnamefont{Holzapfel}},\ }%
  \bibfield{journal}{%
  \bibinfo {journal} {Appl. Phys. Lett.}\ }%
  \textbf{\bibinfo {volume} {95}},\ \bibinfo {pages} {192501} (\bibinfo {year}
  {2009})\BibitemShut{NoStop}%
\bibitem{24}%
  \BibitemOpen
  \bibfield{author}{%
  \bibinfo {author} {\bibfnamefont{K.}~\bibnamefont{Iida}}, \bibinfo {author}
  {\bibfnamefont{J.}~\bibnamefont{H{\"a}nisch}}, \bibinfo {author}
  {\bibfnamefont{T.}~\bibnamefont{Thersleff}}, \bibinfo {author}
  {\bibfnamefont{F.}~\bibnamefont{Kurth}}, \bibinfo {author}
  {\bibfnamefont{M.}~\bibnamefont{Kidszun}}, \bibinfo {author}
  {\bibfnamefont{S.}~\bibnamefont{Haindl}}, \bibinfo {author}
  {\bibfnamefont{R.}~\bibnamefont{H{\"u}hne}}, \bibinfo {author}
  {\bibfnamefont{L.}~\bibnamefont{Schultz}},\ and\ \bibinfo {author}
  {\bibfnamefont{B.}~\bibnamefont{Holzapfel}},\ }%
  \bibfield{journal}{%
  \bibinfo {journal} {Phys. Rev. B}\ }%
  \textbf{\bibinfo {volume} {81}},\ \bibinfo {pages} {100507(R)} (\bibinfo
  {year} {2010})\BibitemShut{NoStop}%
\bibitem{39}%
  \BibitemOpen
  \bibfield{author}{%
  \bibinfo {author} {\bibfnamefont{T.}~\bibnamefont{Katase}}, \bibinfo {author}
  {\bibfnamefont{Y.}~\bibnamefont{Ishimaru}}, \bibinfo {author}
  {\bibfnamefont{A.}~\bibnamefont{Tsukamoto}}, \bibinfo {author}
  {\bibfnamefont{H.}~\bibnamefont{Hiramatsu}}, \bibinfo {author}
  {\bibfnamefont{T.}~\bibnamefont{Kamiya}}, \bibinfo {author}
  {\bibfnamefont{K.}~\bibnamefont{Tanabe}},\ and\ \bibinfo {author}
  {\bibfnamefont{H.}~\bibnamefont{Hosono}},\ }%
  \bibfield{journal}{%
  \bibinfo {journal} {Appl. Phys. Lett.}\ }%
  \textbf{\bibinfo {volume} {96}},\ \bibinfo {pages} {142507} (\bibinfo {year}
  {2010})\BibitemShut{NoStop}%
\bibitem{51}%
  \BibitemOpen
  \bibfield{author}{%
  \bibinfo {author} {\bibfnamefont{J.}~\bibnamefont{H{\"a}nisch}}, \bibinfo
  {author} {\bibfnamefont{K.}~\bibnamefont{Iida}}, \bibinfo {author}
  {\bibfnamefont{F.}~\bibnamefont{Kurth}}, \bibinfo {author}
  {\bibfnamefont{A.}~\bibnamefont{Kauffmann}}, \bibinfo {author}
  {\bibfnamefont{M.}~\bibnamefont{Kidszun}}, \bibinfo {author}
  {\bibfnamefont{S.}~\bibnamefont{Haindl}}, \bibinfo {author}
  {\bibfnamefont{T.}~\bibnamefont{Thersleff}}, \bibinfo {author}
  {\bibfnamefont{J.}~\bibnamefont{Freudenberger}}, \bibinfo {author}
  {\bibfnamefont{L.}~\bibnamefont{Schultz}},\ and\ \bibinfo {author}
  {\bibfnamefont{B.}~\bibnamefont{Holzapfel}},\ }%
  \bibinfo {journal} {unpublished}\BibitemShut{NoStop}%
\bibitem{09}%
  \BibitemOpen
\bibfield{journal}{%
    }%
  \bibfield{author}{%
  \bibinfo {author} {\bibfnamefont{S.}~\bibnamefont{Lee}}, \bibinfo {author}
  {\bibfnamefont{J.}~\bibnamefont{Jiang}}, \bibinfo {author}
  {\bibfnamefont{J.~D.}\ \bibnamefont{Weiss}}, \bibinfo {author}
  {\bibfnamefont{C.~M.}\ \bibnamefont{Folkman}}, \bibinfo {author}
  {\bibfnamefont{C.~W.}\ \bibnamefont{Bark}}, \bibinfo {author}
  {\bibfnamefont{C.}~\bibnamefont{Tarantini}}, \bibinfo {author}
  {\bibfnamefont{A.}~\bibnamefont{Xu}}, \bibinfo {author}
  {\bibfnamefont{D.}~\bibnamefont{Abraimov}}, \bibinfo {author}
  {\bibfnamefont{A.}~\bibnamefont{Polyanskii}}, \bibinfo {author}
  {\bibfnamefont{C.~T.}\ \bibnamefont{Nelson}}, \bibinfo {author}
  {\bibfnamefont{Y.}~\bibnamefont{Zhang}}, \bibinfo {author}
  {\bibfnamefont{S.~H.}\ \bibnamefont{Baek}}, \bibinfo {author}
  {\bibfnamefont{H.~W.}\ \bibnamefont{Jang}}, \bibinfo {author}
  {\bibfnamefont{A.}~\bibnamefont{Yamamoto}}, \bibinfo {author}
  {\bibfnamefont{F.}~\bibnamefont{Kametani}}, \bibinfo {author}
  {\bibfnamefont{X.~Q.}\ \bibnamefont{Pan}}, \bibinfo {author}
  {\bibfnamefont{E.~E.}\ \bibnamefont{Hellstrom}}, \bibinfo {author}
  {\bibfnamefont{A.}~\bibnamefont{Gurevich}}, \bibinfo {author}
  {\bibfnamefont{C.~B.}\ \bibnamefont{Eom}},\ and\ \bibinfo {author}
  {\bibfnamefont{D.~C.}\ \bibnamefont{Larbalestier}},\ }%
  \bibfield{journal}{%
  \bibinfo {journal} {Appl. Phys. Lett.}\ }%
  \textbf{\bibinfo {volume} {95}},\ \bibinfo {pages} {212505} (\bibinfo {year}
  {2009})\BibitemShut{NoStop}%
\bibitem{14}%
  \BibitemOpen
  \bibfield{author}{%
  \bibinfo {author} {\bibfnamefont{A.}~\bibnamefont{Yamamoto}}, \bibinfo
  {author} {\bibfnamefont{J.}~\bibnamefont{Jaroszynski}}, \bibinfo {author}
  {\bibfnamefont{C.}~\bibnamefont{Tarantini}}, \bibinfo {author}
  {\bibfnamefont{L.}~\bibnamefont{Balicas}}, \bibinfo {author}
  {\bibfnamefont{J.}~\bibnamefont{Jiang}}, \bibinfo {author}
  {\bibfnamefont{A.}~\bibnamefont{Gurevich}},\ and\ \bibinfo {author}
  {\bibfnamefont{D.C.}\ \bibnamefont{Larbalestier}},\ }%
  \bibfield{journal}{%
  \bibinfo {journal} {Appl. Phys. Lett.}\ }%
  \textbf{\bibinfo {volume} {94}},\ \bibinfo {pages} {062511} (\bibinfo {year}
  {2009})\BibitemShut{NoStop}%
\bibitem{10}%
  \BibitemOpen
  \bibfield{author}{%
  \bibinfo {author} {\bibfnamefont{S.}~\bibnamefont{Lee}}, \bibinfo {author}
  {\bibfnamefont{J.}~\bibnamefont{Jiang}}, \bibinfo {author}
  {\bibfnamefont{C.~T.}\ \bibnamefont{Nelson}}, \bibinfo {author}
  {\bibfnamefont{C.~W.}\ \bibnamefont{Bark}}, \bibinfo {author}
  {\bibfnamefont{J.~D.}\ \bibnamefont{Weiss}}, \bibinfo {author}
  {\bibfnamefont{C.}~\bibnamefont{Tarantini}}, \bibinfo {author}
  {\bibfnamefont{H.~W.}\ \bibnamefont{Jang}}, \bibinfo {author}
  {\bibfnamefont{C.~M.}\ \bibnamefont{Folkman}}, \bibinfo {author}
  {\bibfnamefont{S.~H.}\ \bibnamefont{Baek}}, \bibinfo {author}
  {\bibfnamefont{A.}~\bibnamefont{Polyanskii}}, \bibinfo {author}
  {\bibfnamefont{D.}~\bibnamefont{Abraimov}}, \bibinfo {author}
  {\bibfnamefont{A.}~\bibnamefont{Yamamoto}}, \bibinfo {author}
  {\bibfnamefont{Y.}~\bibnamefont{Zhang}}, \bibinfo {author}
  {\bibfnamefont{X.~Q.}\ \bibnamefont{Pan}}, \bibinfo {author}
  {\bibfnamefont{E.~E.}\ \bibnamefont{Hellstrom}}, \bibinfo {author}
  {\bibfnamefont{D.~C.}\ \bibnamefont{Larbalestier}},\ and\ \bibinfo {author}
  {\bibfnamefont{C.~B.}\ \bibnamefont{Eom}},\ }%
  \bibfield{journal}{%
  \bibinfo {journal} {Nat. Mater.}\ }%
  \textbf{\bibinfo {volume} {9}},\ \bibinfo {pages} {397} (\bibinfo {year}
  {2010})\BibitemShut{NoStop}%
\bibitem{43}%
  \BibitemOpen
  \bibfield{author}{%
  \bibinfo {author} {\bibfnamefont{C.}~\bibnamefont{Tarantini}}, \bibinfo
  {author} {\bibfnamefont{S.}~\bibnamefont{Lee}}, \bibinfo {author}
  {\bibfnamefont{Y.}~\bibnamefont{Zhang}}, \bibinfo {author}
  {\bibfnamefont{J.}~\bibnamefont{Jiang}}, \bibinfo {author}
  {\bibfnamefont{C.~W.}\ \bibnamefont{Bark}}, \bibinfo {author}
  {\bibfnamefont{J.~D.}\ \bibnamefont{Weiss}}, \bibinfo {author}
  {\bibfnamefont{A.}~\bibnamefont{Polyanskii}}, \bibinfo {author}
  {\bibfnamefont{C.~T.}\ \bibnamefont{Nelson}}, \bibinfo {author}
  {\bibfnamefont{H.~W.}\ \bibnamefont{Jang}}, \bibinfo {author}
  {\bibfnamefont{C.~M.}\ \bibnamefont{Folkman}}, \bibinfo {author}
  {\bibfnamefont{S.~H.}\ \bibnamefont{Baek}}, \bibinfo {author}
  {\bibfnamefont{X.~Q.}\ \bibnamefont{Pan}}, \bibinfo {author}
  {\bibfnamefont{A.}~\bibnamefont{Gurevich}}, \bibinfo {author}
  {\bibfnamefont{E.~E.}\ \bibnamefont{Hellstrom}}, \bibinfo {author}
  {\bibfnamefont{C.~B.}\ \bibnamefont{Eom}},\ and\ \bibinfo {author}
  {\bibfnamefont{D.~C.}\ \bibnamefont{Larbalestier}},\ }%
  \bibfield{journal}{%
  \bibinfo {journal} {Appl. Phys. Lett.}\ }%
  \textbf{\bibinfo {volume} {96}},\ \bibinfo {pages} {142510} (\bibinfo {year}
  {2010})\BibitemShut{NoStop}%
\bibitem{32}%
  \BibitemOpen
  \bibfield{author}{%
  \bibinfo {author} {\bibfnamefont{E.~J.}\ \bibnamefont{Kramer}},\ }%
  \bibfield{journal}{%
  \bibinfo {journal} {J. Appl. Phys.}\ }%
  \textbf{\bibinfo {volume} {44}},\ \bibinfo {pages} {1360} (\bibinfo {year}
  {1973})\BibitemShut{NoStop}%
\bibitem{33}%
  \BibitemOpen
  \bibfield{author}{%
  \bibinfo {author} {\bibfnamefont{R.}~\bibnamefont{Prozorov}}, \bibinfo
  {author} {\bibfnamefont{M.~A.}\ \bibnamefont{Tanatar}}, \bibinfo {author}
  {\bibfnamefont{R.~T.}\ \bibnamefont{Gordon}}, \bibinfo {author}
  {\bibfnamefont{C.}~\bibnamefont{Martin}}, \bibinfo {author}
  {\bibfnamefont{H.}~\bibnamefont{Kim}}, \bibinfo {author}
  {\bibfnamefont{V.~G.}\ \bibnamefont{Kogan}}, \bibinfo {author}
  {\bibfnamefont{N.}~\bibnamefont{Ni}}, \bibinfo {author}
  {\bibfnamefont{M.~E.}\ \bibnamefont{Tillman}}, \bibinfo {author}
  {\bibfnamefont{S.~L.}\ \bibnamefont{Bud$'$ko}},\ and\ \bibinfo {author}
  {\bibfnamefont{P.~C.}\ \bibnamefont{Canfield}},\ }%
  \bibfield{journal}{%
  \bibinfo {journal} {Physica C}\ }%
  \textbf{\bibinfo {volume} {469}},\ \bibinfo {pages} {582} (\bibinfo {year}
  {2009})\BibitemShut{NoStop}%
\bibitem{40}%
  \BibitemOpen
  \bibfield{author}{%
  \bibinfo {author} {\bibfnamefont{G.}~\bibnamefont{Blatter}}, \bibinfo
  {author} {\bibfnamefont{M.~V.}\ \bibnamefont{Feigel$'$man}}, \bibinfo
  {author} {\bibfnamefont{V.~B.}\ \bibnamefont{Geshkenbein}}, \bibinfo {author}
  {\bibfnamefont{A.~I.}\ \bibnamefont{Larkin}},\ and\ \bibinfo {author}
  {\bibfnamefont{V.~M.}\ \bibnamefont{Vinokur}},\ }%
  \bibfield{journal}{%
  \bibinfo {journal} {Rev. Mod. Phys.}\ }%
  \textbf{\bibinfo {volume} {66}},\ \bibinfo {pages} {1125} (\bibinfo {year}
  {1994})\BibitemShut{NoStop}%
\end{thebibliography}
\end{document}